\documentclass[onecolumn]{aa} 

\usepackage{graphicx}
\usepackage{txfonts}
\begin{document} 

   \title{Cosmic-ray flux predictions and observations for and with Metis on board Solar Orbiter}

   

   \author{C. Grimani \inst{1,2}
          \and V. Andretta \inst{3} 
          \and P. Chioetto \inst{4,5}
          \and V. Da Deppo \inst{4} 
          \and M. Fabi\inst{1,2}
          \and S. Gissot \inst{6} 
          \and G. Naletto \inst{4,7}
          \and A. Persici\inst{1}
          \and C. Plainaki \inst{8}
          \and M. Romoli \inst{9,10}
          \and F. Sabbatini \inst{11}
          \and D. Spadaro \inst{12}
          \and M. Stangalini \inst{8}
          \and D. Telloni \inst{13}
          \and M. Uslenghi \inst{14}
          \and E. Antonucci	\inst{13}
          \and A. Bemporad \inst{13}
          \and G. Capobianco \inst{13}
          \and G. Capuano \inst{15}
          \and M. Casti	\inst{16}
          \and Y. De Leo \inst{15,17}
          \and S. Fineschi \inst{13}
          \and F. Frassati \inst{13}
          \and F. Frassetto \inst{4}
          \and P. Heinzel \inst{18}
          \and G. Jerse \inst{19}
          \and F. Landini \inst{13}
          \and A. Liberatore \inst{13}
          \and E. Magli \inst{20}
          \and M. Messerotti \inst{19}
          \and D. Moses \inst{21}
          \and G. Nicolini \inst{13}
          \and M. Pancrazzi \inst{13}
          \and M.~G. Pelizzo \inst{22}
          \and P. Romano \inst{12}
          \and C. Sasso \inst{3}
          \and U. Schühle \inst{17}
          \and A. Slemer \inst{4}
          \and T. Straus \inst{3}
          \and R. Susino \inst{13}
          \and L. Teriaca \inst{17}
          \and C.~A. Volpicelli \inst{13}
          \and J. L. Freiherr von Forstner \inst{23}
          \and P. Zuppella \inst{4}
          }

   \institute{DiSPeA, University of Urbino Carlo Bo, Urbino (PU), Italy\\
        \email{catia.grimani@uniurb.it}
    \and{INFN, Florence, Italy}
    \and{INAF – Astronomical Observatory of Capodimonte, Naples, Italy}
    \and{CNR – IFN, Via Trasea 7, 35131, Padova, Italy}
    \and{CISAS, Centro di Ateneo di Studi e Attivit\`a Spaziali "Giuseppe Colombo", via Venezia 15, 35131 Padova, Italy}
    \and{Solar-Terrestrial Centre of Excellence – SIDC, Royal Observatory of Belgium, Ringlaan -3- Av. Circulaire, 1180 Brussels, Belgium}
    \and{Dip. di Fisica e Astronomia “Galileo Galilei”, Università di Padova, Via G. Marzolo, 8, 35131, Padova Italy}
    \and{ASI – Italian Space Agency, Via del Politecnico snc, 00133 Rome, Italy}
    \and{University of Florence, Physics and Astronomy Department, Largo E. Fermi 2, 50125 Florence, Italy}
    \and{INAF Associated Scientist, Italy}
    \and{Alma Mater University, Bologna, Italy}
    \and{INAF – Astrophysical Observatory of Catania, Italy}
    \and{INAF – Astrophysical Observatory of Torino, Italy}
    \and{INAF – Institute for Space Astrophysics and Cosmic Physics, Milan, Italy}
    \and{University of Catania, Physics and Astronomy Department "Ettore Majorana", via S. Sofia 78, 95123 Catania, Italy}
    \and{Catholic University @ NASA – GSFC, Maryland, USA}
    \and{MPS, G{\"o}ttingen, Germany}
    \and{Academy of Science of the Czech Republic}
    \and{INAF - Astrophysical Observatory of Trieste, Italy}
    \and{Politecnico di Torino, Italy}
    \and{NASA HQ, Washington, DC, USA}
    \and{National Research Council of Italy and Institute for Electronics, Information Engineering and Telecommunications, University of Padova, Department of Information Engineering via Gradenigo, 6B, 35131 Padova}
    \and{Institute of Experimental and Applied Physics, Kiel University,24118 Kiel, Germany}
    }     
   \date{}

 
  \abstract
   {The Metis coronagraph is one of the remote sensing instruments hosted on board the ESA/NASA Solar Orbiter mission.  Metis is devoted to carry out the first simultaneous  imaging of the solar corona in both visible light (VL) and ultraviolet (UV). High-energy particles can penetrate spacecraft materials and may limit the performance of the on-board instruments. A study of the galactic cosmic-ray (GCR) tracks observed in the first VL images gathered by Metis during the commissioning phase
   is presented here. A similar analysis is planned for the UV channel.}
   {A prediction of the GCR flux up to  hundreds of GeV is made here for  the first part of the Solar Orbiter mission to study the Metis coronagraph performance.}
   {GCR model predictions  are compared to observations  gathered  on board Solar Orbiter  by the High-Energy Telescope in the range  10~MeV-100~MeV in the summer 2020 and with previous measurements. Estimated cosmic-ray fluxes above 70 MeV n$^{-1}$ have been also parameterized and used  for Monte Carlo simulations aiming at reproducing the cosmic-ray track observations in the Metis coronagraph VL images. The same parameterizations can also be used to study the performance of other detectors.}
   {By comparing observations of cosmic-ray tracks in the Metis VL images with FLUKA Monte Carlo simulations of cosmic-ray interactions in the VL detector, it is found that cosmic rays fire only a fraction of the order of 10$^{-4}$ of the whole image pixel sample. 
   It is also found that the overall efficiency for cosmic-ray identification in the Metis VL images is approximately equal to the contribution of Z$\geqslant$2 GCR particles. A similar study will be carried out during the overall  Solar Orbiter mission duration for instrument diagnostics and to verify whether the Metis data and Monte Carlo simulations would allow for long-term GCR proton flux monitoring.}
   {}


   \maketitle
%
\section{Introduction}
The ESA/NASA Solar Orbiter mission \citep{a&asoloscience,a&asolomission} was launched on  February 10, 2020 at 5:03 AM CET from Cape Canaveral (Florida, USA)  during the solar minimum period between the solar cycle 24 and the solar cycle 25, characterized by a positive polarity period of the global solar magnetic field (GSMF). Four \textit{in situ} and six remote sensing instruments were placed on board the spacecraft (S/C) to study how the Sun generates and controls the heliosphere \citep{a&asoloscience}. The S/C will reach a minimum distance  from the Sun of 0.28 AU and a maximum inclination about the solar equator of 33 degrees during the mission lifetime. Galactic cosmic-ray (GCR) and solar energetic particle (SEP) interactions in the material surrounding the sensitive part of the instruments generate secondary particles that limit the performance  of the detectors on board space missions  \citep[see for instance][]{solwind14, apj1, apj2, grimani20}. In particular, GCRs and SEPs affect the quality of  visible light (VL, in the range 580-640 nm) and ultraviolet  (UV, in a $\simeq$ 10 nm band  around the 121.6 nm H$\textsc{i}$ Lyman-$\alpha$ line) images 
of the Metis coronagraph \citep{a&ametis, andre}. 
Metis, mounted on an external panel of the Solar Orbiter S/C, is shielded by a minimum of 1.2 g cm$^{-2}$ of material and consequently  is traversed by protons and nucleons  
with energies $>$ 10 MeV n$^{-1}$. An analysis of cosmic-ray signatures in the VL  images is presented in this work.
Cosmic-ray observations gathered on board Solar Orbiter between 10 MeV and 100 MeV with the High-Energy Telescope (HET) of the Energetic Particle Detector (EPD) instrument \citep{a&aepd,johan21, sishet, wimmer21} and  model predictions of GCR energy spectra above 70 MeV n$^{-1}$  are also considered in the following for Monte Carlo simulations \citep{flukacern1,flukacern2, flair} aiming at reproducing and validating the tracks of cosmic rays observed in the Metis VL images. A similar analysis will be carried out for the UV channel in the future.

GCRs consist approximately of 98\% of protons and helium nuclei, 1\% electrons and 1\% nuclei with Z$\geq$3, where percentages are in particle numbers to the total number \citep{papini96}. This work is focused on protons and $^4$He nuclei since these particles constitute the majority of the cosmic-ray bulk and the measurements carried out on board Solar Orbiter allow us to optimize the Monte Carlo simulations by selecting proper input fluxes within the range of predictions. Rare particles such as heavy nuclei and electrons will be considered in the future when data will be published by magnetic spectrometer experiments \citep{aguilar18} up to high energies for the period under study. At present time uncertainties on the predictions of the energy spectra of rare particles would be higher than their contribution to the observations carried out with the Metis coronagraph. 
The overall GCR flux shows time, energy, charge and space  modulation in the inner heliosphere
\citep{grimani20}, an aspect of particular importance also in the context of planetary Space Weather and Solar System exploration \citep{plainaki16, plainaki20}. In particular, the GCR flux shows modulations associated with the 11-year solar cycle and the 22-year polarity reversal of the GSMF. These long-term modulations
are ascribable to the particle propagation against the outward motion
of the solar wind and embedded magnetic field \citep{balogh}. 
Particles that at the interstellar medium have energies below tens of MeV are convected
outward before reaching the inner heliosphere, while interplanetary  and interstellar GCR
energy spectra do not show any significant difference above tens of GeV \citep{pogorelov}.

Adiabatic cooling represents the dominant energy loss of cosmic rays observed in the inner heliosphere in the energy range 10-100 MeV.
At these energies cosmic rays can be considered an expanding adiabatically isolated gas of particles that does work  on the surrounding medium thus reducing its internal energy. 
Above tens of MeV cosmic rays diffuse, scatter and
drift through the solar wind and across inhomogeneities of the interplanetary magnetic field
\citep{jokipii}.
The total residence time of cosmic rays in the  heliosphere is strongly energy-dependent
and varies
from hundreds
of days at 100 MeV n$^{-1}$ to tens of days above 1 GeV \citep{pogorelov}. 
Positively charged GCRs undergo  a drift process (see for instance Grimani et al. \citeyear{grim04, grim07}) during negative polarity periods of the GSMF (when the Sun magnetic field lines exit from the Sun South Pole) propagating mainly sunward from the equator along  the heliospheric current sheet. Negatively charged particles would suffer the same drift process during positive polarity periods (epochs during which the Sun magnetic field lines exit from the Sun North Pole). As a result,
during negative (positive) polarity periods the positively (negatively) charged cosmic rays are more modulated than during positive (negative) polarity epochs.    

The spatial dependence of the GCR proton flux was studied with different S/C gathering data simultaneously between tens of MeV and GeV energies. The intensity of these particles was observed to decrease
by a few percent with decreasing radial distance from the Sun and to vary $\ll$ 1\% with                     
increasing heliolatitude with gradients depending on the GSMF polarity epoch \citep{desimone}.                                                                                                                                    



Solar Orbiter was launched during a positive polarity period of the Sun, therefore the drift process played no role in modulating the overall GCR flux in 2020. The same will apply until the next polarity change expected at the maximum of the solar cycle 25 between 2024 and 2025 \citep{sing19}.

This manuscript is arranged as follows:      
in Sect. 2  the radial and latitudinal gradients of the cosmic-ray flux in the inner heliosphere along the Solar Orbiter orbit  are discussed. In Sect. 3 the GCR flux predictions and observations are presented for the summer 2020.
In Sect. 4 the Metis coronagraph is briefly described. In Sect. 5 the S/C on-board algorithm for cosmic-ray track selection in the Metis VL images is illustrated. In Sect. 6 a viewer developed for the Metis cosmic-ray matrices is presented. In Sect. 7 the GCR  observations with Metis are reported. 
Finally,  in Sect. 8 the simulations of the Metis VL detector carried out with the FLUKA Monte Carlo program \citep{flukacern1, flukacern2, flair} are discussed and compared to  observations. The capability of the Metis VL detector to play the role of a cosmic-ray monitor is also evaluated.
\section{Galactic cosmic-ray flux radial and latitudinal gradients in the inner heliosphere}

The GCR proton flux gradients with radial distance from the Sun and heliolatitude were
 studied with simultaneous observations gathered by S/C moving along  different orbits in order to disentangle the role of space and time variations.  
The majority of space instruments are flown  near Earth. In particular, those missions sent to space during the first part of the solar cycle 23 and the second part of the solar cycle 24 experienced the same polarity of the GSMF as Solar Orbiter. Ulysses, launched on October 6, 1990, was placed in an elliptical orbit  around the Sun inclined at 80.2 degrees to the solar equator \citep{ulysses}. The mission was switched off in June 2009.
The Kiel Electron Telescope \citep{ket} on board Ulysses
measured electrons, protons and helium nuclei from MeV to GeV energies.
On June 15, 2006 the Pamela (Payload for Antimatter Matter
Exploration and Light-nuclei Astrophysics; Picozza et al. \citeyear{picozza07}) satellite  
 was launched to collect proton, nucleus and electron data near Earth
above 70 MeV n$^{-1}$ during a period characterized by a negative polarity of the GSMF and a minimum solar modulation. During the same time Ulysses covered a distance from the Sun ranging between 2.3 and 5.3 AU.
The comparison between
Ulysses and Pamela overlapping measurements 
indicated
 for the proton flux in the rigidity
interval 1.6-1.8 GV (0.92-1.09 GeV; corresponding approximately to the median energy of the GCR spectrum at solar minimum) 
a radial intensity gradient of 2.7$\pm$0.2\%/AU
and a latitudinal gradient of $-$0.024$\pm$0.005\%/degree \citep{desimone}.
Positive (negative) latitudinal gradients are observed during positive (negative) polarity periods.
Also the Experiment 6 (E6) on board Helios-A and Helios-B  provided  ion data from 4 to several hundreds of  MeV n$^{-1}$ \citep{heliosAB, helios6}.
The Helios-A and Helios-B S/C were launched on December 10, 1974 and January 15, 1976 during a positive polarity epoch and  sent into ecliptic orbits of  190 and 185 days
periods
around the Sun. The orbits perihelia were 0.3095 AU and 0.290 AU, respectively. The aphelia were approximately 1 AU. As a result, the  Helios data are  representative of the cosmic-ray bulk variations that will be experienced by Solar Orbiter that will also reach maximum distances from the Sun of about 1 AU.
In the recent paper by \citet{helios6} the Helios proton data radial gradients of the GCR flux were found of 6.6$\pm$4\% above 50 MeV 
and 2$\pm$2.5\% between 250 and 700 MeV
between 0.4 and 1 AU: these results agree with those by Pamela/Ulysses within errors.
In conclusion, the variations of the GCR proton dominated flux along the Solar Orbiter orbit are expected to be of a few \% at most and, consequently, it is plausible to assume that
models for cosmic-ray modulation developed on the basis of observations  gathered near Earth will also apply to Solar Orbiter. On the other hand, the Metis data will allow to verify this assumption in the unexplored region of tens of degrees above the solar equator. Analogously, even though no SEP data were gathered up to present time, it will be likely possible to study the evolution of SEP events near the Sun above the solar equator for the first time also with Metis in addition to the dedicated instruments flown on board Solar Orbiter.

\section{Galactic cosmic-ray energy spectra in the summer 2020 for Solar Orbiter}

The solar activity was observed to modulate the near-Earth cosmic-ray integral flux during the last three solar cycles from solar minimum (e.g. years 2009 and 2019) through solar 
maximum (years 1989-1991) above 70 MeV n$^{-1}$  by approximately a factor of four. This period of time is considered here since the solar modulation was studied in the majority of cases with data gathered in space\footnote{\url{http://cosmicrays.oulu.fi/phi/Phi_mon.txt}}. The above estimate was carried out with the  
Gleeson and Axford model \citep[G\&A;][]{glax68}. The same model is also adopted here to estimate the  GCR flux  modulation  by considering  
the observed  average monthly sunspot number a proxy of the solar activity\footnote{Data used here are publicly available at \url{http://www.sidc.be/silso/datafiles}} \citep{silso}. 
This model  allows us to estimate  the cosmic-ray intensity in the inner heliosphere by assuming an interstellar energy spectrum and
 a solar modulation parameter ($\phi$) that basically represents the energy loss of cosmic rays propagating from  the interstellar medium
to  the point of observations. During GSMF positive polarity epochs, the G\&A model is found to reproduce well the  GCR measurements at 1 AU in the energy 
range  from tens of MeV to hundreds of GeV \citep{grim07}. It is pointed out that different values of the solar modulation parameter are estimated if different GCR energy spectra at the interstellar medium are considered in the model. Voyager 1 measured the interstellar spectra 
of ions and electrons below 1 GeV \citep{stone2013}. However, the solar modulation parameter\footnote{see \url{http://cosmicrays.oulu.fi/phi/Phi_mon.txt}} adopted in this work \citep{uso11, uso17} was estimated according to the \citet{burger2000} interstellar proton spectrum. Since the solar modulation parameter and 
the interstellar particle spectra must be properly associated, the interstellar proton spectrum by \citet{burger2000} is also considered here. Unfortunately, no $^4$He interstellar spectrum is reported in this last paper and, consequently, we adopted the  estimate by 
\citet{Shikaze2007154} that was inferred from the balloon-borne BESS experiment data \citep{bess14} gathered during different solar modulation and solar polarity periods. This choice is not expected  to affect the overall simulation outcomes by more than a few \% since the cosmic-ray bulk is dominated  by protons. A similar approach was considered for the LISA Pathfinder mission orbiting around the first Lagrange point during the years 2016-2017 \citep{armano2016,armano18}. It was shown 
in \cite{mattia} that the  integral proton flux predictions carried out with the G\&A model for LISA Pathfinder differ by less than 10\% from the Alpha Magnetic Spectrometer (AMS-02) experiment data \citep{ams01, aguilar18} gathered on the Space Station during the Bartels Rotation 2491 \citep{grim19}. For the present work, it is also possible to benefit of the proton and helium differential flux measurements of the EPD/HET
instrument flying on board  Solar Orbiter and gathering data in the energy range below 100 MeV 
to further reduce the uncertainty on the GCR flux predictions while waiting for the AMS-02 data to be published for the years after 2017 up to TeV energies. 
In Fig. 1 the average monthly sunspot number observed during the solar cycles 23 and 24 is shown. Minimum, average and maximum predictions of the sunspot number are inferred from the Marshall Space Flight Center website\footnote{\url{https://solarscience.msfc.nasa.gov/predict.shtml}} and appear as dotted and dashed lines in the same figure. The launch of the Solar Orbiter S/C is also indicated. For the first 2-3 years of  the Solar Orbiter mission, it is plausible to expect a minimum to low solar modulation analogously to the years 1996-1997 at the beginning of the solar cycle 23 during a positive polarity epoch.

\begin{figure}[ht]
\begin{center}
\centering \includegraphics[width=0.6\hsize]{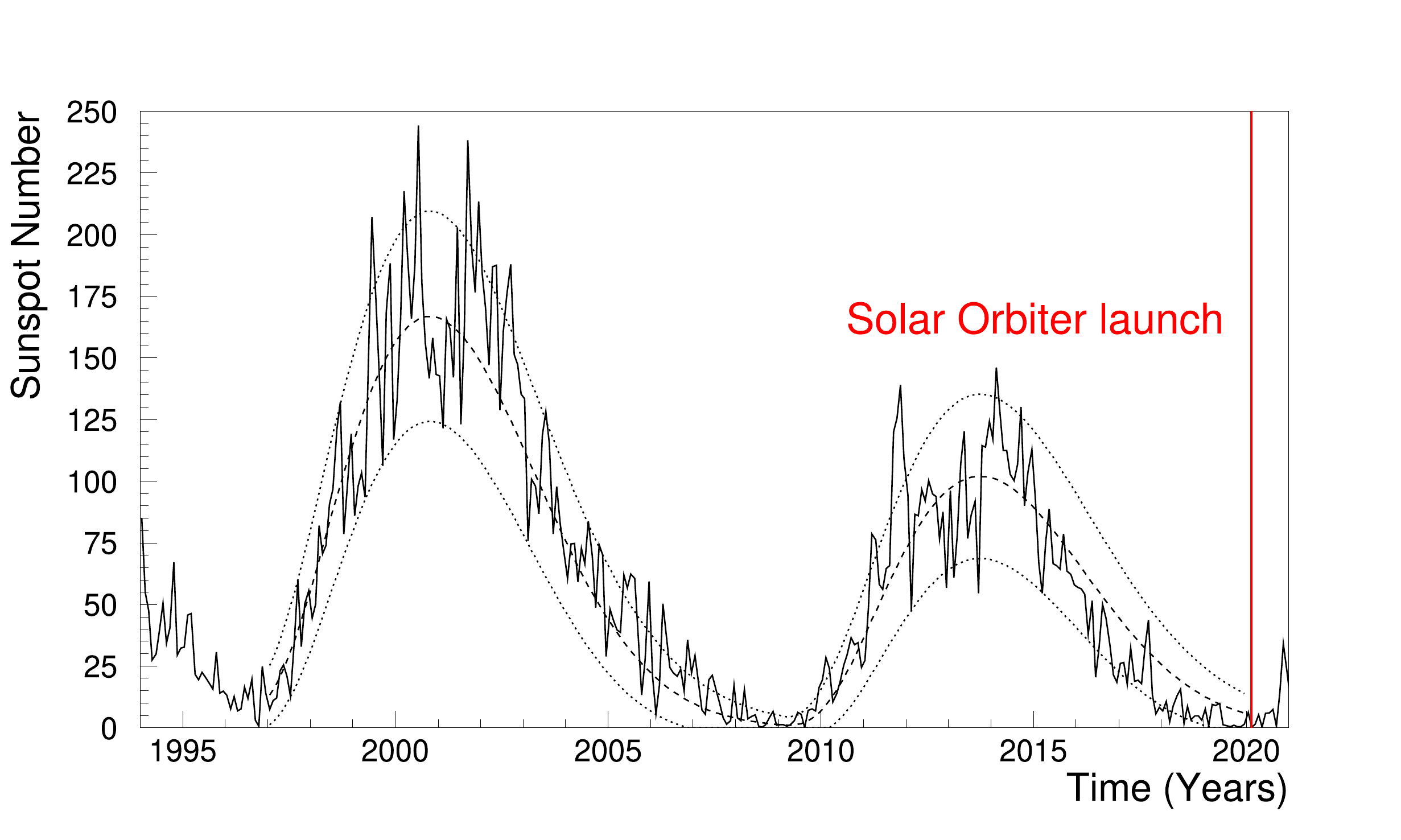}
\end{center}
\caption{\label{ssn} Average monthly sunspot number observed since 1994 \citep{silso}.}
\end{figure}

\begin{figure}[ht]
\begin{center}
\centering \includegraphics[width=0.6\hsize]{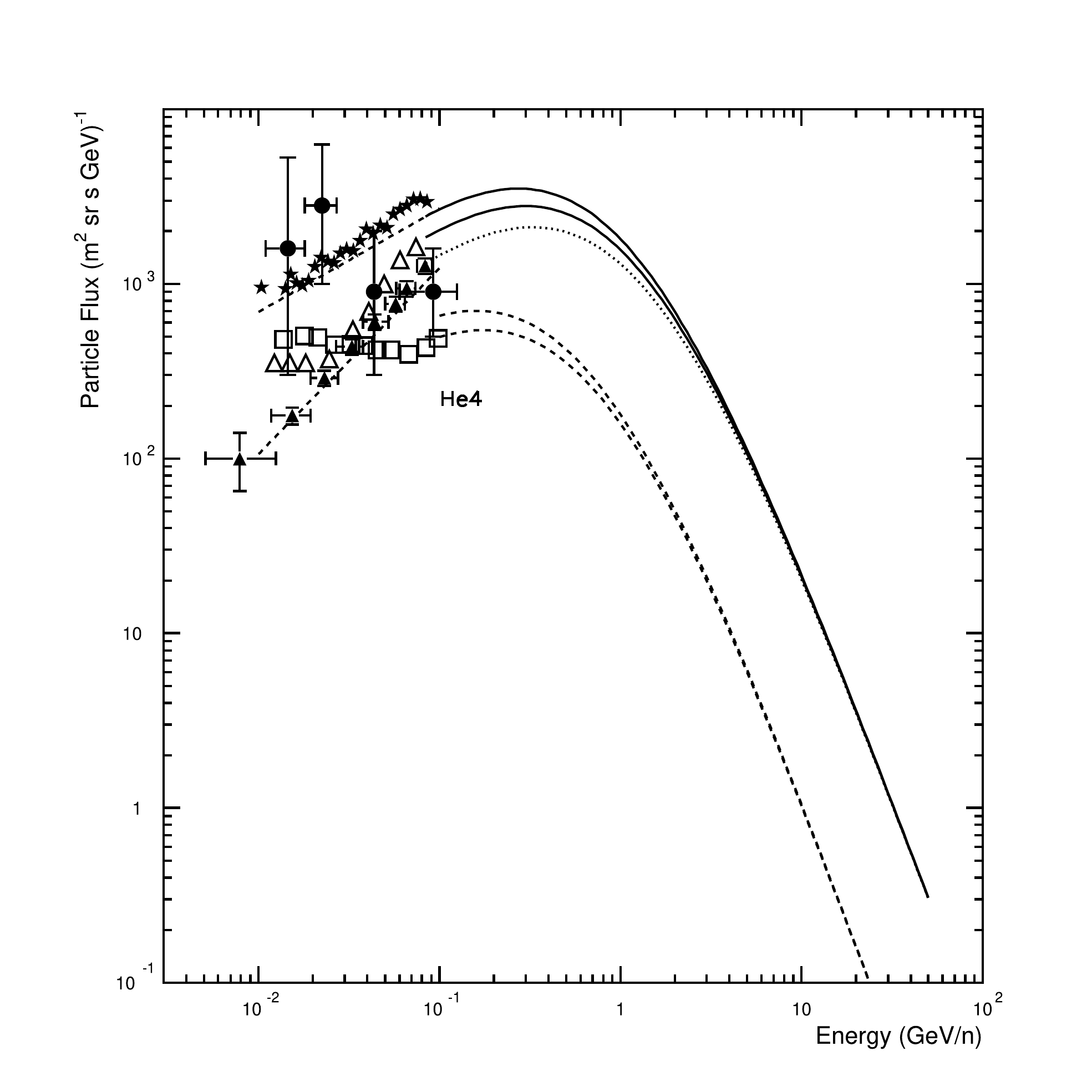}
\end{center}
\caption{\label{metpred} Predictions of proton and $^4$He nucleus fluxes during the summer 2020 for the Solar Orbiter mission. The top and bottom continuous (protons) and dashed ($^4$He nuclei) curves are obtained with the G\&A model above 70 MeV(/n) for a solar modulation parameter of 300 and 340 MV/c, respectively.  The dotted curve represents the lower limit to the predictions of the proton spectrum for the same period, representing the Pamela experiment data gathered in 2008 during a period of solar activity similar to that of Solar Orbiter but during a  negative GSMF epoch. Data are reported in \citet[][solid dots]{stone64}; \citet[][open triangles]{logachev} and \citet[][solid triangles]{helios6}. The solid stars and the open squares indicate the EPD/HET measurements gathered in the summer 2020 on board Solar Orbiter for protons and helium nuclei, respectively.}
\end{figure}

\begin{table}
\centering
\caption{\label{table1} Average monthly sunspot number observed after the Solar Orbiter launch.}
\begin{tabular}{ll}
\hline
\hline
   & Monthly sunspot number   \\
\hline
February 2020& 0.2     \\
March 2020& 1.5      \\
April 2020& 5.2      \\
May 2020& 0.2      \\
June 2020&  5.8    \\
July 2020& 6.1      \\
August 2020& 7.5     \\
September 2020& 0.6      \\
October 2020& 14.4      \\
November 2020& 34.0      \\
December 2020& 21.8      \\
January 2021& 10.4      \\
\hline
\end{tabular}
\end{table}
The solar activity has been very low since 2019, when a yearly sunspot number of 3.6 was observed, and slightly increased in 2020.
The average monthly sunspot number observed after the Solar Orbiter launch is also reported in Table~\ref{table1}.  
It is focused in particular on the period June-July 2020, when the EPD/HET data were released for the first time and the average sunspot number was 6.0.
The solar modulation parameter for this period  is reasonably assumed to range  between  300 MV c$^{-1}$  and  340 MV c$^{-1}$. In particular, a solar modulation of 340 MV c$^{-1}$ corresponds to the average solar modulation parameter observed during periods of a similar solar activity observed in the past years, while 300 MV c$^{-1}$ is considered a lower limit for $\phi$ corresponding to the GCR flux modulation that seems to better agree with the EPD/HET data below 100 MeV reported as black stars in Fig.~\ref{metpred} after rebinning. The uncertainty on the integral flux predictions is 16\% only because of the small solar modulation variation between June and July 2020. The EPD/HET data are compared below 100 MeV with measurements carried out by other experiments during similar conditions of solar modulation and GSMF polarity  (see Table~\ref{table2}).  
The estimated proton and $^4$He  energy spectra appear also in Fig.~\ref{metpred} with continuous lines for protons and with dashed lines for $^4$He. 

We are aware that in-flight calibration of the EPD/HET instrument is still ongoing and that, in particular, the proton fluxes above 10~MeV are overestimated by a factor of up to two as discussed by \citet{wimmer21}. However, in the following we consider $\phi$ = 300 MV c$^{-1}$ because this parameter modulation value allows us to reconcile our galactic cosmic-ray flux predictions with HET measurements within normalization uncertainty and because it appears reasonable with respect to 320 MV c$^{-1}$ observed at the time of the LISA Pathfinder mission in mid-2017 when the solar modulation was higher than in 2020 \citep{apj1}.

The particle energy spectra above 70 MeV have been parameterized as follows (see for details Grimani et al.  \citeyear{grimani20}):

\begin{equation}
F(E)= A\ (E+b)^{-\alpha}\ E^{\beta}  \ \ \ {\rm particles/(m^2\ sr\ s\ GeV/n),}
\label{equation1}
\end{equation}

\noindent where $E$ is the particle kinetic energy per nucleon in GeV/n.
The parameters $A$, $b$, $\alpha$ and $\beta$  for  minimum (m) and
maximum (M) predictions are shown in Table~\ref{table3}, along with the lower limit (ll) for the estimated proton flux in 2020 set by the Pamela experiment observations carried out in 2008 during a period of low solar modulation and negative GSMF polarity \citep{pamela11}.  

\begin{table*}
\caption{\label{table2} Solar modulation and GSMF polarity for the datasets reported in Fig.~\ref{metpred}\tablefootmark{a}.}
\centering                          
\begin{tabular}{llll} 
\hline\hline
Experiment  & Time &  Solar activity ($\phi$) &  Solar polarity  \\
  &  &  MV/c  &    \\
\hline

HET/EPD  & June-July 2020 & low ($\phi$=300-340) & $>$0 \\
Helios/E6 &1974-1978 & low ($\phi$=424-544) & $>$0 \\
IMP-8 & 14-16 May 1976 & low ($\phi$=434) & $>$0 \\
Near-Earth satellite & 13 December 1961 & average ($\phi$=698) & $>$0 \\
\hline
\end{tabular} 
\tablefoot{
\tablefoottext{a}{Data are reported in \citet{stone64,logachev,helios6}.}}\\
\end{table*}

\begin{table*}  
\caption{\label{table3} Parameterizations of proton and helium energy spectra in June-July 2020
above 70 MeV n$^{-1}$\tablefootmark{b}.}    
\centering
\begin{tabular}{@{}*{13}{l}}
\hline  
\hline
 &  $A_m$  &  $b_m$ &  $\alpha_m$  & $\beta_m$ &  $A_M$ &  $b_M$ &  $\alpha_M$ &  $\beta_{M}$  &   $A_{ll}$ &  $b_{ll}$ &  $\alpha_{ll}$ &  $\beta_{ll}$ \\ 
\hline 
p  & 18000 & 0.95 & 3.66 & 0.87 & 18000 & 0.875 & 3.66 & 0.87 & 18000 & 1.05 & 3.66 & 0.87\\                                                                          
He & 850  & 0.58  & 3.68 & 0.85 & 850 & 0.53 & 3.68 & 0.85 & & & &\\
\hline                                                                                                                                                                 
\end{tabular}
\tablefoot{
\tablefoottext{b}{Maximum (M; top continuous and dashed lines in Fig.~\ref{metpred}), minimum (m; bottom continuous and dashed lines in Fig.~\ref{metpred}) and lower limit proton predictions 
(ll; dotted line in Fig.~\ref{metpred}). The units of the parameters $A$ and $b$ are particles/(m$^2$ sr s GeV/n$^{-\alpha +\beta +1}$) and GeV/n, respectively, while the spectral indices $\alpha$ and $\beta$ are pure numbers.}}
\end{table*}

\section{The Metis coronagraph and visible light image characteristics}
Metis is a solar coronagraph performing for the first time simultaneous imaging of the off-limb solar corona in both visible light and 
ultraviolet \citep{a&ametis}. The main design constraints for the Metis coronagraph were associated with the aim of obtaining the major scientific return with VL and UV observations of the corona while suffering large thermal variations (up to about 400~$^{\circ}$C in the region of the inverted external occulter) along the Solar Orbiter orbit. Total mass ($<$ 24.55 kg) and power consumption ($<$ 28 W) were also limited as imposed to interplanetary missions. Moreover, the Metis design was optimized to achieve a sensitivity to observe the weak corona from 1.7 R$_{\odot}$ through 9 R$_{\odot}$ by maintaining a contrast ratio lower than 10$^{-9}$ and pointing the Sun center within one arcmin. 
With an innovative occultation design and a unique combination of mirrors and filters, Metis images the VL corona in the range 580-640 nm (corresponding to yellow-orange light) and the UV H$\textsc{i}$ Lyman-$\alpha$ line at 121.6 nm. The VL channel includes a polarimeter that allows to observe the linearly polarized component of the K corona. The minimum spatial resolution at perihelion is 2000 km for VL images and the   temporal resolution is of up to 1 second.    
The VL detector consists of a VL camera with an active CMOS (CMOSIS ISPHI Rev. B developed by  CMOSIS Imaging Sensors, now AMS, Belgium) sensor segmented in 4.1943 $\times$ 10$^6$ pixels. Each pixel has dimensions of 10$\mu$m$\times$10$\mu$m$\times$4.5$\mu$m \citep{a&ametis} accounting for a pixel and detector  geometrical factors of 401 $\mu$m$^2$ sr  and 17 cm$^2$ sr, respectively \citep{sullivan}.
\section{The Metis algorithm for cosmic-ray detection}
When the VL and UV detectors work in analog mode, several frames are co-added to obtain a single image to improve the signal-to-noise (S/N) ratio \citep{andre}. 
The detection and removal of GCRs and eventually of solar energetic particles, is carried out on board by the Metis processing unit. The primary goal of the particle track removal process is to improve the quality of the images. As a byproduct, the algorithm also provides the location and number of particle tracks in each image which, in principle, can be correlated with the GCR integral flux over the image integration time. No particle identification is allowed. Other instruments such as the extreme UV imager SOHO-EIT \citep{soho-eit} and the coronagraph STEREO-SECCHI \citep{secchi} reported a  number of pixels fired  by GCRs per second per square centimeter  differing by almost two orders of magnitude \citep{andre}. As recalled above, this evidence is not ascribable to a varying solar activity.
The mentioned observations had to result from particle interactions in the S/C material surrounding the instruments.
Therefore, Monte Carlo simulations are needed to correlate the particle tracks observed in the Metis coronagraph images to the 
incident flux of GCRs.

The temporal noise observed in the CMOS pixel signals includes both the readout noise and the statistical noise associated with photons and dark current fluctuations. In principle, this temporal noise can be expressed as follows:

\begin{equation}                        
\sigma^2= A+B\times p, 
\label{equation2} 
\end{equation}                                                                                                                                                                  

\noindent where $p$ is the pixel value and A and B are coefficients that can be fine tuned on the basis of the characterization of the VL detector. 
For each series of N images with exposure times of T seconds each, for each image $i$ of the sequence, starting from $i$=2 and for each pixel value $p$, the difference  

\begin{equation}
\delta p = abs (p_i - p_{i-1}) 
\label{equation3}
\end{equation}

\noindent is calculated. By defining $p_m$=min($p_i$, $p_{i-1}$), the related noise is then estimated as indicated in Eq.~2 with p=p$_m$.


In the case $\delta$$p$$^2$ $>$ k$_{th}$ $\sigma^2$ both pixel values are replaced with $p_m$. This process is repeated for each pixel of each image for $i$=2,..., N. 
The parameter $k_{th}$ is also user defined and determines the detection threshold. 

\begin{figure}[ht]
\begin{center}
\centering \includegraphics[width=0.6\hsize]{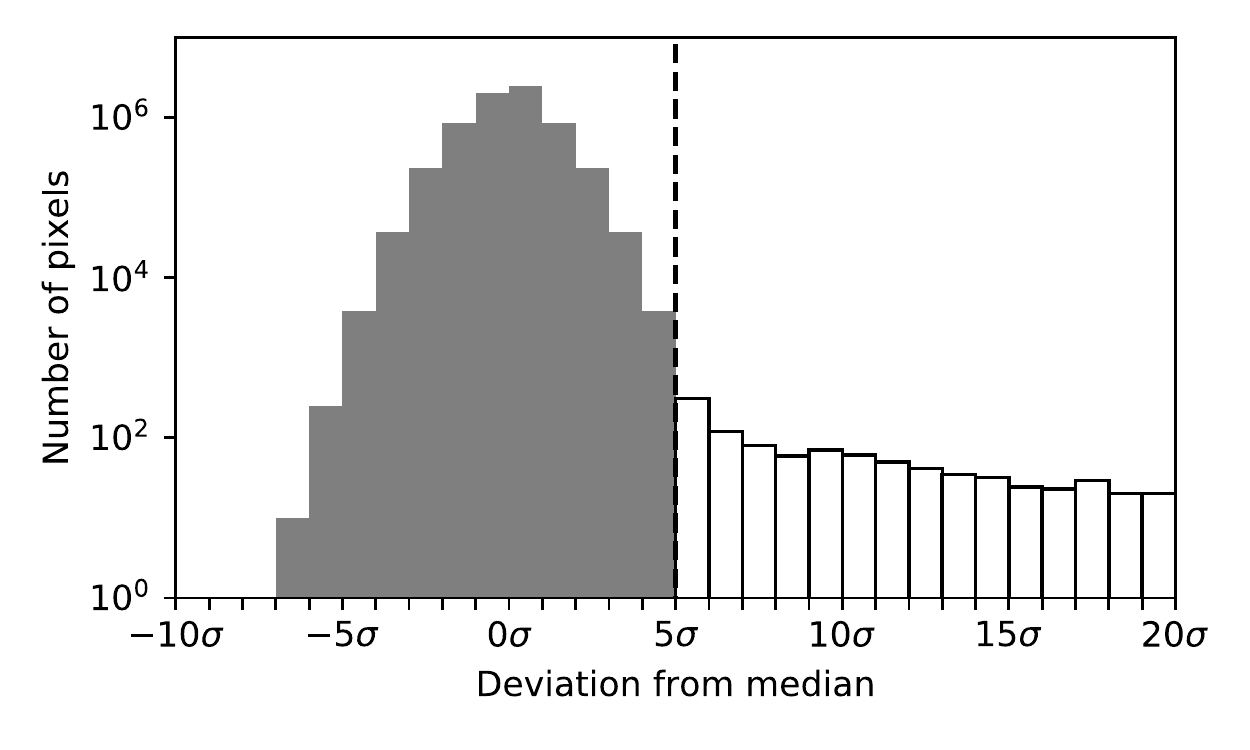}
\end{center}
\caption{\label{cor1} STEREO-B COR 1 data gathered in 2007 and reported in \cite{andre}. The dark area is associated with VL photon signals and dark current, while the white area is ascribable to the passage of high-energy particles in the VL detector.}
\end{figure}

As an example, the STEREO-B COR1 data are shown in Fig.~\ref{cor1} \citep[see][and references therein]{andre}.  
Fig.~\ref{cor1} shows the approximately Gaussian distributions of pixel signals generated by photons and noise represented by the dark area while cosmic-ray signals are  associated with the white area. The threshold is set at 5 $\sigma$ from median.

During the commissioning phase of the Solar Orbiter mission the parameters A and B for Metis were set to 40000 and 0, respectively. Despite the possibility to improve the algorithm performance for cosmic-ray search in the future, very encouraging results have been obtained in the analysis presented here.



\section{The APViewer for cosmic-ray track analysis in the Metis VL images}

\begin{figure}[ht]
\centering \includegraphics[width=0.6\hsize]{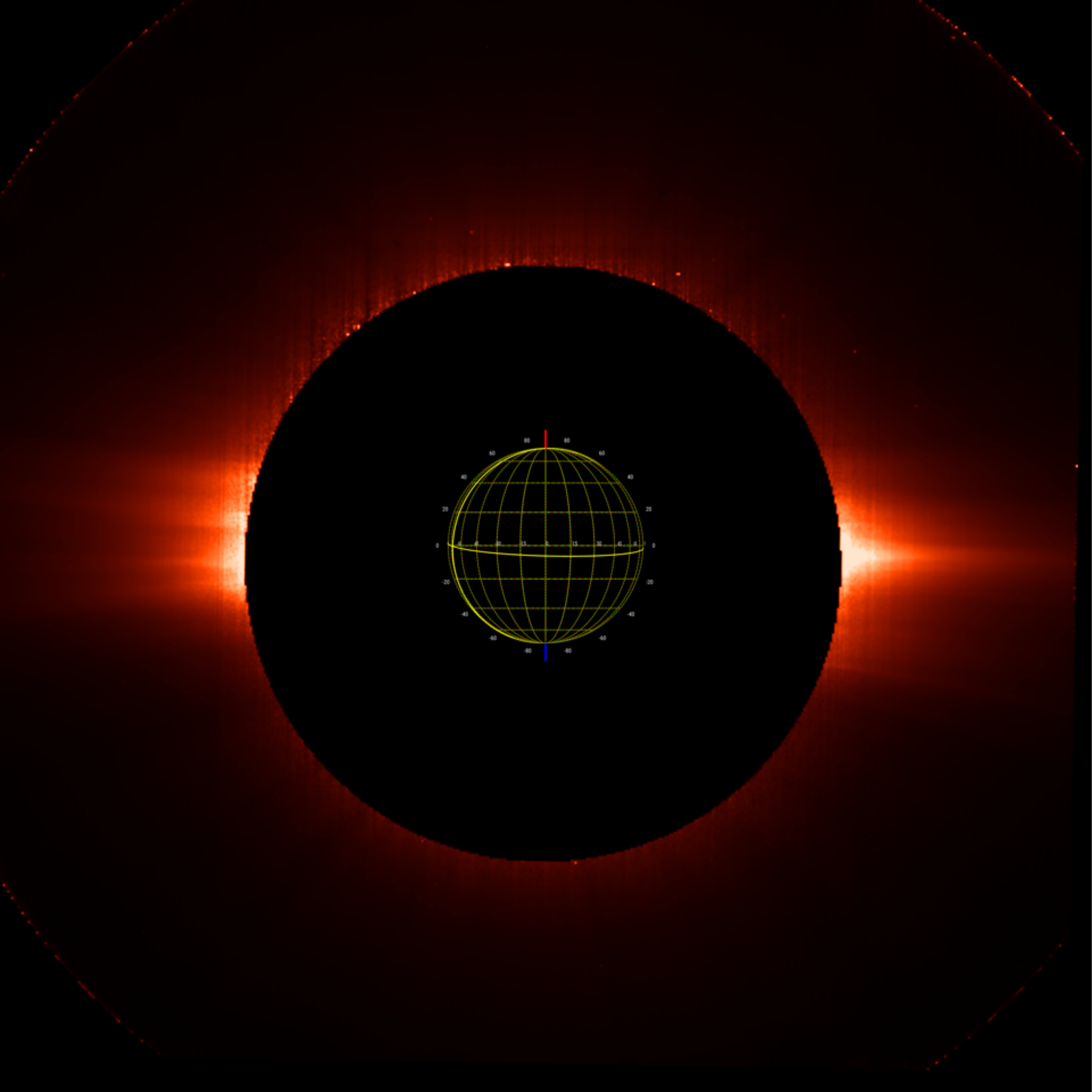}
\caption{\label{vlgreen} Polarized brightness image of the solar corona in VL (June 21, 2020). The vertical orientation is North-South according to the Solar axis. The image was obtained with an exposure time of 30 s at a distance of 0.5 AU from the Sun.}
\end{figure}

\begin{figure*}
\begin{center}
\centering \includegraphics[width=0.6\hsize]{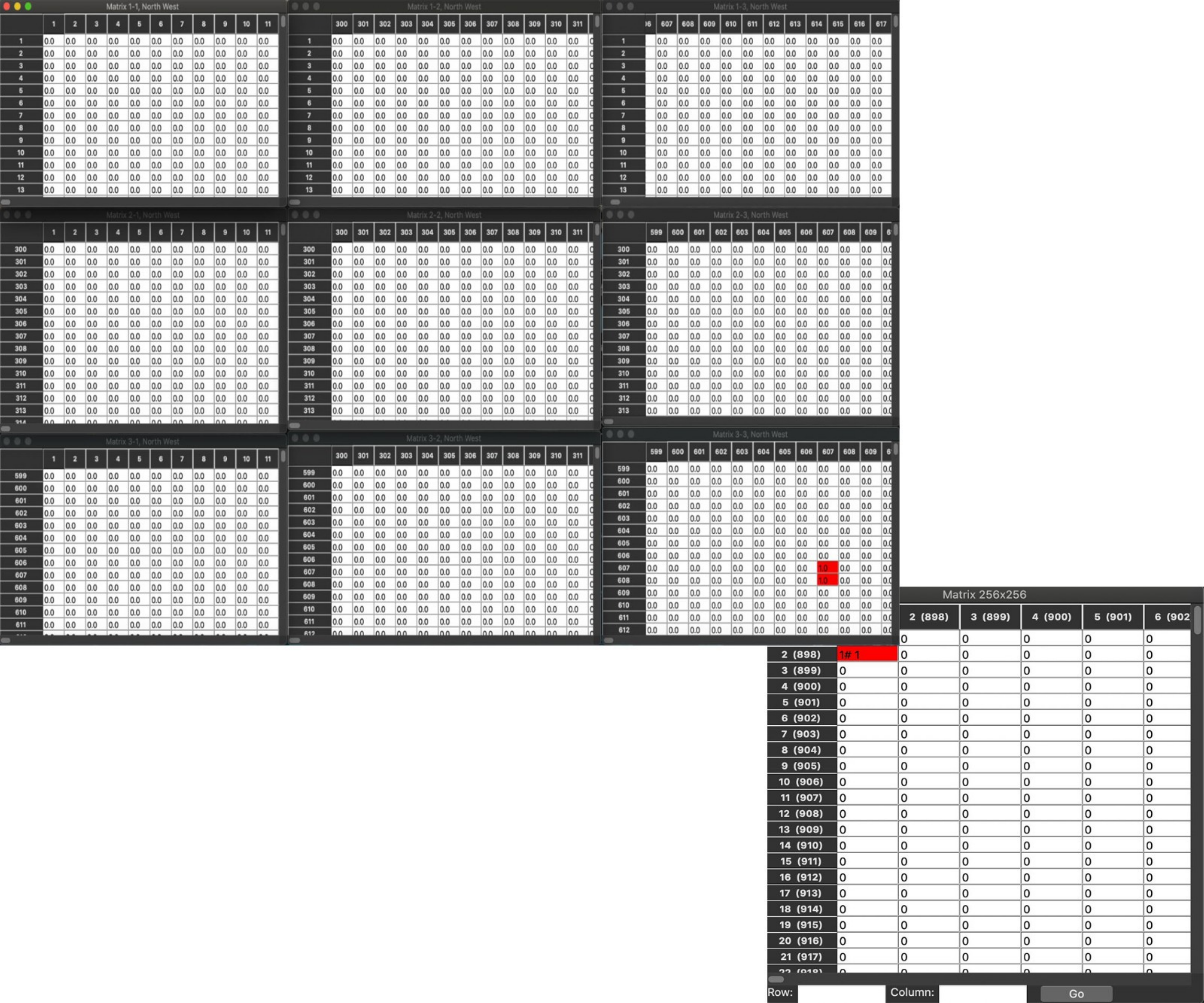}
\end{center}
\caption{\label{apvcorner} Example of cosmic-ray matrix corner window opening with the APViewer.}
\end{figure*}

A Metis VL image  of the solar corona is shown in Fig.~\ref{vlgreen}. After removing the cosmic-ray tracks, the 
  on-board algorithm described in the previous section allows to report the fired pixels  in 2048$\times$2048 matrices called \textit{cosmic-ray matrices}. 
The algorithm cannot discriminate between intense signals generated  by   cosmic rays  from noisy pixels 
and  bright sources that lie in  the detector field of view. However, external bright sources and noisy pixels 
would be possibly found  in the same matrix cells in more than one image of each set of images, while cosmic rays would hit different pixels in each image. As a result,
the superposition of several images allows to increase the statistics for GCR analysis and to improve the S/N ratio. 
In the cosmic-ray matrices the pixel contents correspond to the number of times that, from the pixel-to-pixel comparison among each image of each set of images, 
the pixel content was changed to p$_m$.
For instance, in this work, where a case study  of four sets of four images (N=4) is presented, 
the pixel contents are 0, 1, 2 or 3.
The format of the images is FITS.
A dedicated viewer, APviewer, was developed for Metis in Python programming language for the visual analysis of the particle tracks in the cosmic-ray matrices \citep{apviewer}.  The APViewer allows for an easier visualization and search of the cosmic-ray tracks. Furthermore, it provides an efficient solution to a series of problems that other tools present. In particular, the widely used viewer FV\footnote{ \url{https://heasarc.gsfc.nasa.gov/docs/software/ftools/fv/}} for images in FITS format does not visually differentiate pixels with distinct values and does not provide a zoomed-in view of the region surrounding an identified pattern of fired pixels. To this purpose, the APViewer performs a coloring of the pixels according to their values ($>$ 0) and allows the user to choose the window size when looking for a specific cosmic-ray track in order to guarantee an accurate view of the surrounding context.
In the specific case of VL images, the main window of the APViewer presents a 256x256 matrix that constitutes the central inner portion of the original 2048x2048 matrix. This reduced matrix provides an overview of the location of the pixels traversed by cosmic rays. The contents of the border cells of the reduced matrix are displayed as: $a\#b$, 
where $a$ corresponds to the sum of all the fired pixels (i.e., those with values greater than 0) in the row/column portion of the corresponding side of the matrix, whereas $b$ represents the sum of the pixels with value equal to 1, thus returning the number of pixels traversed by cosmic rays in that same row/column hidden parts.
In order to guarantee a straightforward  search of cosmic-ray tracks, the main window includes two text-boxes where the user can type  the row and column number of the pixel he wishes to look at in more detail. It is worthwhile to point out that fired pixels can also be analyzed through sub-windows that can be opened starting from the 
reduced matrix by means of a left or right click on the border cells. 
These windows vary in size, according to the specific side of the matrix, and in number, depending on whether it is a corner cell or a border one. As an example, in Fig.~\ref{apvcorner} it is shown  how it is possible to open one of the hidden parts of a cosmic-ray matrix starting from the corner of the central inner matrix.  

\section{Analysis of cosmic-ray tracks in the Metis visible light images}

\begin{figure*}
\centering \includegraphics[width=0.6\hsize]{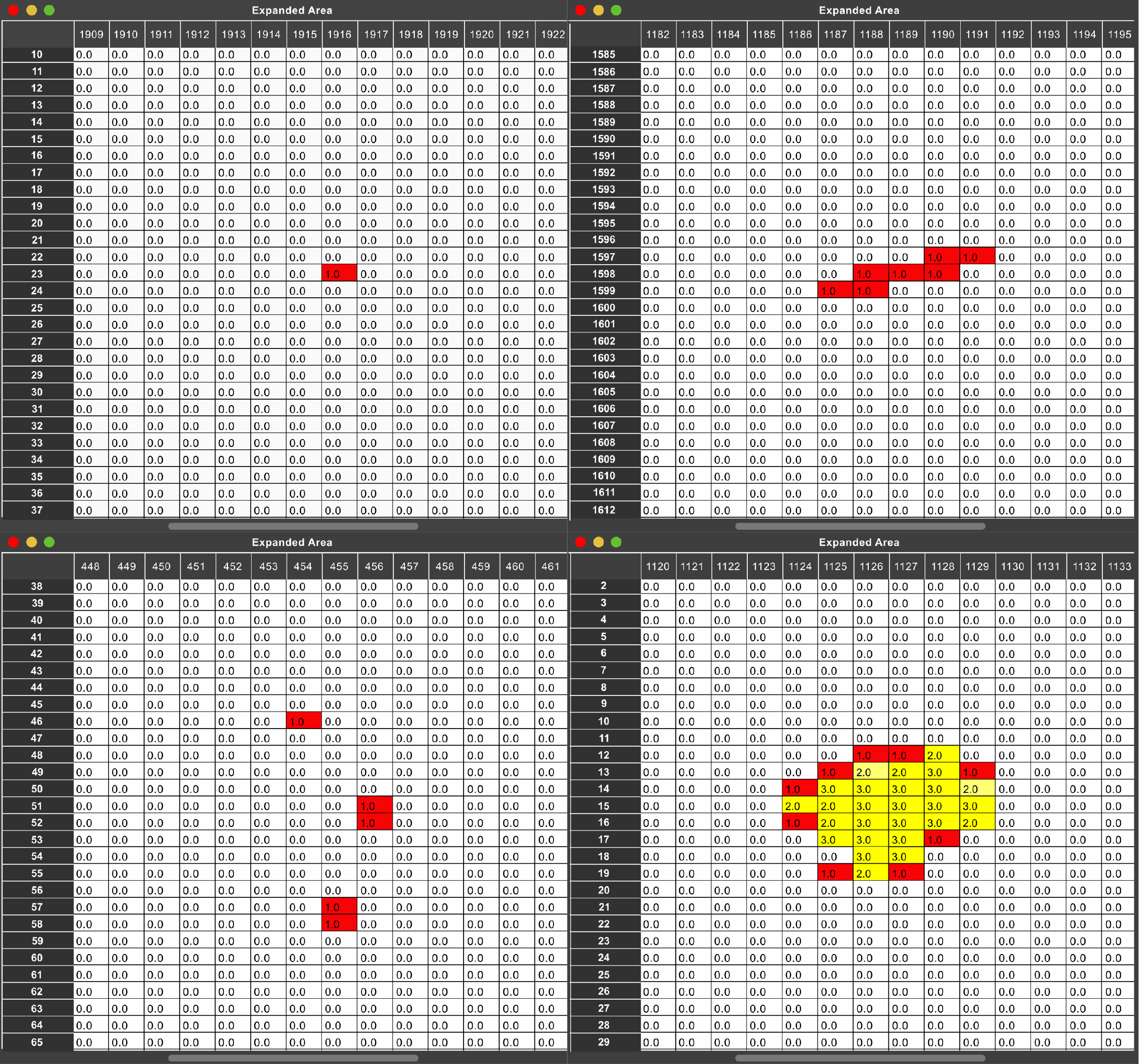}
\caption{\label{apvtracks} Examples of cosmic-ray tracks observations in the Metis VL images (top panels and left bottom panel). A spurious pattern of  hit pixels ascribable to a bright external source such as a star appears in the bottom right panel. }
\end{figure*}

A sample of cosmic-ray  tracks  observed in the Metis VL images is shown in Fig.~\ref{apvtracks}. 
GCR tracks are classified as single hits (top left panel), slant tracks (top right panel) and multiple tracks (bottom left panel). 
Slant tracks are associated with single particles firing more than one pixel. It is  recalled here that GCRs show a spatial isotropic distribution and that each single fired pixel has a geometrical factor and geometrical field of view  of 401 $\mu$m$^2$ sr and 132 degrees, respectively.  Multiple tracks are associated with patterns of fired pixels   characterized by a distance from track centroid smaller than 12 pixels. The image shown in the bottom right panel of Fig.~\ref{apvtracks} is not ascribable to the passage of cosmic rays and represents the effects of a bright external source such as a star. The number of pixels with contents associated to external sources moving through the instrument field of view during the acquisition of images was 10$^{-5}$ of the total pixel sample. 

As recalled above, four sets of four co-added cosmic-ray matrices  characterized by  15 seconds (T) of exposure each, for a total of 60 second exposure time for each set of 
images, were considered here as a case study. These images were taken on May 29, 2020. 
The presence of noisy isolated pixels in the four images was studied by searching for fired pixels appearing in more than one image. No one was found. Conversely, noisy columns were found in two cosmic-ray matrices. These pixels  were not considered for the analysis resulting to be just a fraction of approximately 3$\times10^{-5}$ and 5$\times10^{-5}$, respectively, of the total number of pixels. The efficiency for single pixels being fired  was set with a study of a sample of 23 slant tracks: only 6 pixels out of 106 counted along the particle tracks  were not fired  and consequently the efficiency was set to 0.94$\pm$0.02. This estimate represents a lower limit to the actual efficiency applying to the total bulk of incident particles. The pixel inefficiency depends on particle energy losses in the sensitive part of the detector and, consequently, on the particle species and path length. Straight tracks impacting in the central part of the pixels are certainly characterized by a higher efficiency than slant tracks. Unfortunately, a dedicated beam test to estimate the efficiency of the detector with different particle species and incident directions was not carried out before the mission launch.  
Single pixel  and Metis on-board algorithm efficiencies for cosmic-ray selection must be properly taken into account to reconcile cosmic-ray observations carried out with Metis and the  integral flux of GCRs. 
Single pixel, slant  and multiple tracks have been  counted in each studied VL image.  The results are reported in Table~\ref{table4} where the tracks have been listed in categories according to their topology. The average number of GCR tracks per set of images corresponding to a total of 60 second exposure time was 271$\pm$22.

\begin{table*}
\centering
\caption{\label{table4} Cosmic-ray tracks in the studied sets of Metis VL images. Square corresponds to track patterns of four fired pixels forming a square. The other definitions are intuitive.}
\begin{tabular}{lllllll}
\hline
\hline
 Image Set & Single pixel & Horizontal & Vertical & Slant & Square & Number of tracks \\
\hline
1 & 183 & 20 & 22 & 27 & 3 & 255 \\
2 & 200 & 28 & 35 & 33 & 7 & 303 \\
3 & 186 & 27 & 22 & 27 & 0 & 262 \\
4 & 183 & 33 & 21 & 24 & 4 & 265 \\
\hline
\end{tabular}
\end{table*}

The visual analysis of the Metis VL images did not allow to distinguish between primary and secondary particles generated by incident cosmic rays interacting
in the material surrounding the coronagraph.
Monte Carlo simulations of the VL detector were carried out to this purpose. 

 
\section{Comparison of simulations and cosmic-ray observations in the Metis visible light images}
While single pixels  and slant tracks observed in the Metis VL images are certainly associated with single  particles, the multiple tracks could have been generated by 
incident particle interactions. A toy Monte Carlo allowed to study the random incidence of 300 particles (roughly equivalent to the number of tracks observed in each set of VL images) on the active part of the detector. Ten different runs were carried out and the average minimum distance observed between different particle tracks  was of 5 pixels
($\simeq$ 50 $\mu$m), consistently with multiple track observations.

\begin{figure}[ht]
\begin{center}
\centering \includegraphics[width=0.6\hsize]{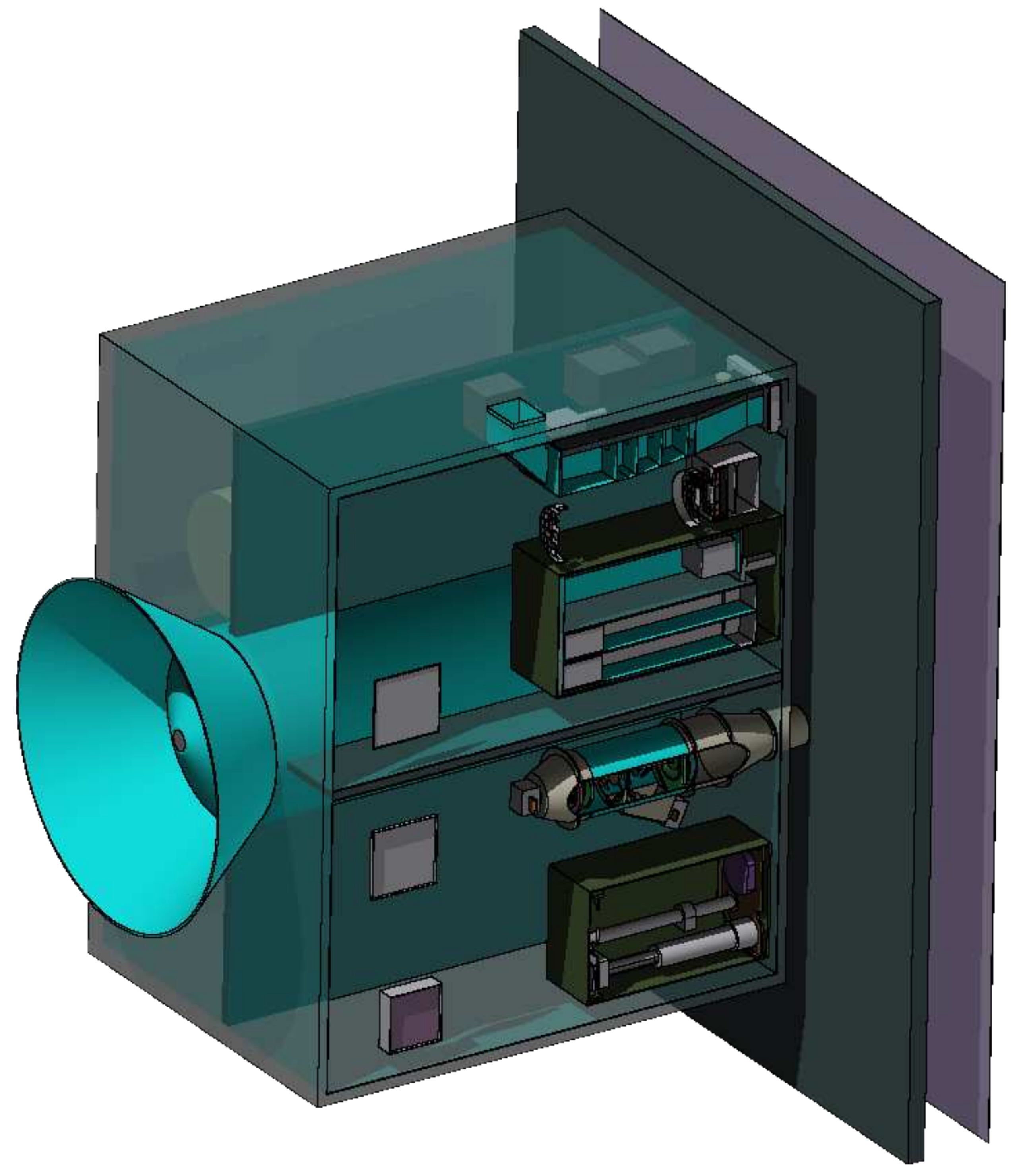}
\end{center}
\caption{\label{flukamodel} Solar Orbiter geometrical model. Remote sensing instruments and electronic boxes are visible.}
\end{figure}

In Fig.~\ref{flukamodel} the geometry of the Solar Orbiter S/C and instruments adopted for the Metis VL detector simulations carried out with the FLUKA Monte Carlo program (version 4.0.1) is shown. The geometry includes the S/C structure, thrusters fuel tanks and the SPICE, EUI, PHI and STIX instruments \citep[][and references therein]{a&asolomission}. Interplanetary and galactic particles traverse from about 1 g cm$^{-2}$ to more than 10 g cm$^{-2}$ of material depending on the particle incidence direction before reaching the Metis VL detector.
Proton and helium fluxes reported in Fig.~\ref{metpred} as top continuous and dashed lines were considered as input GCR energy 
differential fluxes for the summer 2020 above 70 MeV in the simulations. Particles with energies below   70 MeV n$^{-1}$ were found not to give a relevant contribution to cosmic-ray tracks in the VL images. 
The number of incident particles was set in order to reproduce 
the 60 seconds exposure time of the VL detector images. Longer simulations would have been unfeasible because of limited available computational power. 
The simulations returned 276$\pm$17 charged particles crossing the VL detector for incident proton flux only.  This number of tracks is similar to the observations within statistical uncertainties. It is worthwhile to point out that the statistical error is smaller than the combination of the uncertainties on input fluxes of the order of 10\% on the basis of our previous experience with LISA Pathfinder \citep{grim19} and the intrinsic Monte Carlo resolution also of the order of 10\%  \citep{fluka_uncertainty}. While the detector does not allow to distinguish different particle species, the simulations indicate that for primary protons the particles traversing the sensitive part of the detector in particle numbers to the total number down to 1\% in composition are:  80\% protons,  17\% electrons and positrons, 3\% pions. For the helium run particles crossing the sensitive part of the VL detector were ascribable for 30\% of the total bulk of tracks to He$^{4}$ nuclei, 34\% to electrons and positrons, 25\% to protons, 10\% to pions and 1\% to deuterium. It is also found that the number of tracks generated by He$^{4}$ amounts to 30\% of the total sample generated by primary protons. The contribution of nuclei with Z$>$2 was estimated to be 5\% of the total. This last estimate was carried out with a separate set of Monte Carlo simulations with which the relative role of protons, helium and heavy nuclei was set at solar minimum according to \cite{grim1}, since no heavy nucleus measurements were available for the summer 2020. The total number of cosmic-ray tracks in the Metis images is dominated by primary and secondary particles associated with incident protons. It is also noticed that the algorithm and detector efficiencies remove 35\% of the tracks from the images, the same percentage of pixels fired by helium and other nuclei. Finally,  the minimum distance between secondary particles generated by the same incident primary on the S/C ranged between 2 cm and 500 $\mu$m. As a result, the FLUKA simulations 
confirmed the toy Monte Carlo results indicating that multiple tracks in the VL images are produced by different incident particles. 
According to these results, if confirmed during the rest of the mission when the solar modulation is expected to increase and after the next polarity change in 2024-2025, by assuming that no significant variation of the VL instrument sensitivity for GCR detection will be observed, Metis proton dominated data joint with Monte Carlo simulations may allow for the monitoring of long-term variations of GCRs above 70 MeV. 
 When the magnetic spectrometer experiment AMS-02 will provide absolute GCR flux measurements in 2020 for model prediction normalization and with the implementation of the full Solar Orbiter S/C geometry, it will be plausibly possible to reduce the uncertainties on the simulations and, consequently, on the Metis cosmic-ray matrices analysis. Finally, a comparison of the Metis observations with those of other instruments flown on board Solar Orbiter, such as EUI-FSI \citep{a&aeui}, will allow us to further test the outcomes of this work. 

\section{Conclusions}
High-energy particles traverse and interact in the S/C materials of space missions, thus limiting the efficiency of on-board instruments. A dedicated Python viewer was developed to study the cosmic-ray tracks observed in the Metis VL images of the solar corona. It was found that in 60 seconds of exposure time, the number of pixels traversed by cosmic rays is about a fraction of 10$^{-4}$ of the total number of pixels and therefore the quality of the images is not significantly affected by cosmic ray tracks. Monte Carlo simulations of the VL detector are consistent with the number of observed tracks when only primary protons are considered.
The signatures of the passage of particles with charge $>$1 account approximately for the single pixel and on-board algorithm overall efficiencies. Consequently, Metis data and Monte Carlo simulations may allow for the monitoring of proton flux  long-term variations. Future analysis of Metis VL and UV cosmic-ray matrices and comparison with measurements of other instruments such as EUI-FSI will allow to further test our work.

\begin{acknowledgements}
Solar Orbiter is a space mission of international collaboration between ESA and NASA, operated by ESA.  The Metis programme is supported by the Italian Space Agency (ASI) under the contracts to the co-financing National Institute of Astrophysics (INAF): Accordi ASI-INAF N. I-043-10-0 and Addendum N. I-013-12-0/1, Accordo ASI-INAF N.2018-30-HH.0  and under the contracts to the industrial partners OHB Italia SpA, Thales Alenia Space Italia SpA and ALTEC: ASI-TASI N. I-037-11-0 and ASI-ATI N. 2013-057-I.0. Metis was built with hardware contributions from Germany (Bundesministerium für Wirtschaft und Energie (BMWi) through the Deutsches Zentrum für Luft- und Raumfahrt e.V. (DLR)), from the Academy of Science of the Czech Republic (PRODEX) and from ESA. \\
We thank J. Pacheco and J. Von Forstner of the EPD/HET collaboration for useful discussions about cosmic-ray data observations gathered on board Solar Orbiter. We thank also the  PHI and EUI Collaborations for providing useful details about instrument geometries for S/C simulations.
\end{acknowledgements}

%
   \bibliographystyle{aa} 
   \bibliography{biba17feb21} 
%

\end{document}